\documentstyle[11pt]{article}
\begin{document}
\title{Testing Cosmic Censorship in Kerr-like Collapse Situations 
\thanks{This essay received an ``honorable mention'' from the Gravity
Research Foundation for 1997.} \thanks{to be published in Mod. Phys.
Lett. A.}} 
\author{{\bf Wies{\l}aw Rudnicki} \\ {\em Institute of Physics,
Pedagogical University of Rzesz\'ow,} \\ {\em ul. Rejtana 16A,
PL-35-310 Rzesz\'ow, Poland} \\ {\em E-mail: rudnicki@univ.rzeszow.pl}}
\date{}
\maketitle
\begin{abstract}According to the cosmic censorship hypothesis of
Penrose, naked singularities should never occur in realistic collapse
situations. One of the major open problems in this context is the
existence of a naked singularity in the Kerr solution with $|a|>m$; 
this singularity can be interpreted to be the final product of collapse
of a rapidly rotating object. Assuming that certain very general and
physically reasonable conditions hold, we show here, using the global
techniques, that a realistic gravitational collapse of any rotating
object, which develops from a regular initial state, cannot lead to the
formation of a final state resembling the Kerr solution with a naked
singularity. This result supports the validity of the cosmic censorship
hypothesis.  
\end{abstract}

\vspace{6mm}
%\newpage

The cosmic censorship hypothesis put forward by Penrose \cite{p1} is
perhaps the most important unsolved problem in classical general
relativity. This hypothesis concerns spacetime singularities which are
expected to occur, according to the famous Hawking-Penrose theorem
\cite{hp}, as a result of collapse of massive stars. The occurrence of
these singularities implies that classical general relativity breaks
down whenever too large concentrations of mass are present. The
additional problem here is the question of whether or not these
singularities will causally influence any regular parts of spacetime
and hence give rise to breakdowns of predictability in these regions.
This important issue is closely related to the cosmic censorship
hypothesis which asserts that singularities arising in realistic
collapse situations will always be hidden inside an event horizon and
hence shielded from the external view (no naked singularities). Thus,
if this hypothesis is true, the possible breakdowns of predictability
due to singularities might only occur inside black holes, and so one
could ignore this pathology because it will never cause any detectable
effects for observers staying in the external world. Recall also here
that the assumption that naked singularities do {\em not} occur plays
an essential role in proving a number of important theorems in general 
relativity: black hole uniqueness theorem, black hole area theorem, and
positive mass theorem.  

There exist various examples of exact solutions of Einstein's equations
that represent the collapse of a regular initial configuration to a
{\em naked} singularity (see, e.g., \cite{kl} and references cited
therein). All these examples, however, represent special situations
that may never arise in a realistic collapse, and so they cannot,
without proofs of stability, have any bearing on the hypothesis of
Penrose. For instance, most of them (with the notable exception of the
case explored numerically by Shapiro and Teukolsky \cite{st}) deal only
with the collapse of {\em nonrotating} configurations of matter. But as
realistic collapsing stars are expected to be generally rotating, one
should attempt to test cosmic censorship in more general collapse
scenarios, where angular momentum is present as well. One of the known
exact solutions of Einstein's equations which might be relevant to such
a situation is that of Kerr \cite{ke}. This solution depends on two
parameters $m$ and $a$, and represents the exterior gravitational field
of a rotating body with mass $m$ and angular momentum $L=am$, as
measured from infinity in gravitational units $(G=c=1)$. From the point
of view of cosmic censorship, the most interesting feature of the Kerr
solution is the ringlike curvature singularity appearing in the central
part of the solution whenever $m\neq 0$ and $a\neq 0$ --- this
singularity can be interpreted to be the final product of collapse of a
rotating object.  

There is, however, an essential difference between the Kerr singularity
existing in the case $|a|\leq m$ and that occurring as $|a|$ exceeds
the mass $m$. Namely, if $|a|\leq m$, then there exists an event
horizon and the whole ring singularity is always hidden inside this
horizon, whereas if $|a|>m$, there is {\em no} event horizon and the 
singularity is visible for all observers (see \cite{he}, pp. 161-168). 
Accordingly, since many stars may have an angular momentum greater than
the square of their mass (for the Sun, $L\sim m^{2}$), it is of
importance in view of cosmic censorship to know whether a physically
realistic collapse of a rotating object, which starts off from a
regular initial state, can ever lead to the formation of a final 
state resembling the Kerr solution with a naked singularity. Exactly
this question has been raised by Penrose in his original article on
cosmic censorship \cite{p1} as the basic open problem in the context of
his hypothesis. In this letter, we shall present a theorem which provides
an answer to the above question. Our approach is based on the global
techniques; our notation will be the same as that of Hawking and 
Ellis \cite{he}.  

Consider a spacetime $(M,g)$ admitting a weakly asymptotically simple
and empty conformal completion $(\tilde{M},\tilde{g})$ (\cite{he}, p.
225). Such spacetimes are ideally suited to model the collapse of
isolated objects. Let $C$ be a partial Cauchy surface for $(M,g)$; if
the future null infinity ${\cal J}^{+}$ of the completion
$(\tilde{M},\tilde{g})$ is contained in the closure of the future
Cauchy development $D^{+}(C,\tilde{M})$, then $(M,g)$ is said to be
{\em future asymptotically predictable from} $C$ (\cite{he}, p. 310).
Future asymptotic predictability is a precise statement of cosmic
censorship for $(M,g)$, since it assures that there will be no
singularities to the future of an initial data surface which are
naked, i.e. visible from ${\cal J}^{+}$. Now suppose that the spacetime
$(M,g)$ contains a rotating object which undergoes complete
gravitational collapse. Assume also that the collapse starts off from a
{\em regular} initial state. To make this notion precise, we shall
assume that the collapse develops from initial data given on some
partial Cauchy surface $S$ of $(M,g)$ which satisfies the following two
conditions: (1) $I^{-}(S)=D^{-}(S)$; and (2) {\em all null geodesics
generating ${\cal J}^{+}$ intersect} $\bar{D}^{+}(S,\tilde{M})$.
Condition (1) is an obvious requirement ensuring that all the possible
pathologies of $(M,g)$ --- such as naked singularities or, say,
causality violations --- can only occur to the future of the surface
$S$. Condition (2) means that $(M,g)$ is to be {\em partially} future
asymptotically predictable from $S$ as defined by Tipler \cite{t1}. This
requirement is needed in order to exclude the possibility of artificial
breakdowns of future asymptotic predictability due to a poor choice of
the initial surface. An example of such a poor choice of a partial
Cauchy surface may be the surface $t=-(1+x^{2}+y^{2}+z^{2})^{1/2}$ in
Minkowski space.  

In order to consider the question of whether or not the collapse can
lead to any naked singularity, one should first impose on $(M,g)$ a
condition ruling out artificial naked singularities which can easily
be created just by removing points from $J^{+}(S)\cap J^{-}({\cal
J}^{+},\tilde{M})$. A very useful condition of this type have been
proposed in \cite{r1}. The physical justification for this condition is
the idea that essential singularities should always be associated with 
large curvature strengths, which should in turn lead to the focusing of
Jacobi fields along null geodesics approaching the singularities.  Let
$\lambda(t)$ be an affinely parametrized null geodesic, and let $Z_{1}$
and $Z_{2}$ be two linearly independent spacelike vorticity-free Jacobi
fields along $\lambda(t)$. The exterior product of these Jacobi fields
defines a spacelike area element, whose magnitude at the parameter
value $t$ we denote by $A(t)$. If we now introduce the function $z(t)$
defined by $A(t)\equiv z^{2}(t)$, then one can show \cite{t2} that
$z(t)$ satisfies 
\begin{equation}
\frac{d^{2}z}{dt^{2}}+\frac{1}{2}(R_{ab}K^{a}K^{b}+2\sigma^{2})z=0,
\end{equation}
where $K^{a}$ is the tangent vector to $\lambda(t)$ and $\sigma^{2}$ 
characterizes the shear of Jacobi fields along $\lambda(t)$ (\cite{he},
p. 88). Now consider a past (future) endless null geodesic
$\eta:[0,s)\rightarrow M$ which is past (future) incomplete at the
value $s$ of its affine parameter $t$. Assume also that $\eta$
generates an achronal set (e.g., $\eta$ could be a generator of a
Cauchy horizon). The geodesic $\eta$ is said to satisfy the {\em
inextendibility condition} \cite{r1} if there exist a parameter value 
$t_{1}\in (0,s)$ and a solution $z(t)$ to Eq. (1) along $\eta(t)$ with
initial conditions: $z(t_{1})=0$ and $\dot{z}(t_{1})=1$, such that
$\lim_{t\rightarrow s}z(t)=0$. One can show \cite{r1} that if $\eta$
does satisfy this condition, then there is {\em no} reasonable
extension of the spacetime $M$ in which $\eta$ could be continued
beyond the value $s$ of its affine parameter $t$, and this means that
$\eta$ should then approach a genuine singularity at $s$. Thus we can
avoid the occurrence of artificial naked singularities in $(M,g)$ by
assuming that: (3) {\em every past (future) incomplete, past (future)
endless null geodesic generating an achronal subset of $(M,g)$
satisfies the above inextendibility condition}.  

As a matter of fact, one cannot {\em a priori} exclude the existence of
some ``true'' singularities in $(M,g)$, i.e. those not caused only
by the removal of regular points, such that null geodesics approaching
them fail to satisfy the inextendibility condition. It seems, however,
likely that such singularities would not be accompanied by large
curvature strengths, and so one can hope that they might be ignored on
physical grounds. Indeed, it is worth noting here that null geodesics
terminating at the so-called {\em strong curvature singularities}
\cite{t2}, which are often regarded to be the {\em only} physically
reasonable singularities (see, e.g., \cite{tce,k1}), will always
satisfy, just by definition, the above inextendibility condition. It
should be, however, stressed that our condition will hold for a
considerably larger class of singularities than only those of the
strong curvature type. This is because it does not imply any serious
restrictions on the behaviour of the curvature near singularities (in
fact, it could be satisfied for a given geodesic $\eta$ even if the
curvature along $\eta$ would remain bounded), whereas strong curvature
singularities can exist only if the curvature in their neighbourhood
diverges strong enough \cite{ck}. One can therefore expect that the
condition (3) assumed above should involve no essential loss of
generality of our considerations.  

Now suppose that the above mentioned collapsing object was not able to
dissipate enough the angular momentum to form an event horizon and the
final state of collapse is a region $K\subset J^{+}(S)\cap J^{-}({\cal
J}^{+},\tilde{M})$ resembling just the Kerr solution with a naked
singularity. Clearly, the formation of the nakedly singular region $K$ 
must lead to the occurrence of the future Cauchy horizon, $H^{+}(S)$,
of the initial surface $S$ and one can certainly assume that $K\subset
I^{+}[H^{+}(S)]$. To simplify our considerations, we shall also assume
that the formation of the region $K$ is the {\em only} reason for
occurring of the horizon $H^{+}(S)$. That is, we shall assume that the
following condition holds: (4) {\em Every generator of the Cauchy
horizon $H^{+}(S)$ intersects the boundary $\dot{K}$ of the region
$K$.} Of course, the above assumptions do not guarantee that the region
$K$ will be similar in any nontrivial sense just to the Kerr solution
with a naked singularity. But, as is well known (\cite{he}, p. 162),
the most striking feature of this Kerr solution is the presence of {\em
closed timelike curves} passing through {\em every} point of the
spacetime. One can therefore ensure the existence of a clear similarity
between the region $K$ and the naked Kerr solution by imposing the
requirement that the chronology condition of $(M,g)$ fails to be
satisfied everywhere in $K$. To make things precise, we shall thus
define the naked Kerr-like region $K$ as follows $K:=\{x\in
J^{+}(S)\cap J^{-}({\cal J}^{+},\tilde{M})| \, x\in I^{+}(x) \}$.  

We are now in a position to state the main result. 

\vspace{5mm}
{\bf Theorem.} {\em Under the conditions (1) -- (4) stated above, the
naked Kerr-like region $K$ can never form in the spacetime $(M,g)$ if
the following two additional conditions hold: 

$(a)$ $R_{ab}V^{a}V^{b}\geq 0$ for every null vector $V^{a}$ of
$(M,g)$; 

$(b)$ every null geodesic $\lambda$ of $(M,g)$ admits a point at which 
$K_{[a}R_{b]cd[e}K_{f]}K^{c}K^{d}\neq 0$, where $K^{a}$ is the tangent
vector to $\lambda$.} 

\vspace{5mm}
Conditions $(a)$ and $(b)$ are reasonable requirements for {\em any} 
physically realistic model of a classical spacetime. Since they have
been discussed extensively in the literature on the singularity
theorems (see, e.g., \cite{hp,he}), an extended discussion on them
will not be given here. Note only here that condition $(a)$ may be
obtained, using Einstein's equations, from the weak energy condition:
$T_{ab}K^{a}K^{b}\geq 0$ for any timelike vector $K^{a}$, which is
known to be fulfilled by all the observed classical matter fields.
Condition $(b)$ essentially requires that every null geodesic should
encounter some matter or randomly oriented radiation; one can therefore
expect that this condition should always hold in physically realistic
(generic) spacetimes. It should be, however, stressed here that most of
the known exact solutions of Einstein's equations, due to their special
symmetries, {\em do} violate condition $(b)$; for example, it fails to
be satisfied for null geodesics generating the Cauchy horizons in the
Kerr solution. Thus the above theorem does not exclude the possibility
that the naked Kerr-like region $K$ could occur in some highly
symmetric models of collapse of rotating matter. This theorem shows,
however, that the existence of the region $K$ could not be a {\em
stable} property of such models if they would be slightly perturbed
just enough to satisfy condition $(b)$. One can thus expect that the
naked Kerr-like region $K$ should never form in {\em realistic}
collapse situations.  

In this letter, we shall only give the main ideas of the proof of the
above theorem; for the detailed proof we refer the reader to Ref.
\cite{r2}.  

In very brief outline, the proof runs as follows. First, one
establishes that if the assertion of the theorem were false, then there
would have to exist some past endless, past incomplete generator $\eta$
of the Cauchy horizon $H^{+}(S)$ with future endpoint on ${\cal
J}^{+}$. By condition (4) the generator $\eta$ must intersect the
boundary $\dot{K}$ of the Kerr-like region $K$. This enables one to
show that the causal simplicity condition of $(M,g)$ must break down,
due to the chronology violation inside $K$, at some point $p\in \eta
\cap \dot{K}$, such that $E^{-}(p)\neq \dot{J}^{-}(p)$. Using this, one
then constructs a certain sequence $\{\eta_{i}\}$ of achronal null
geodesic segments converging to the generator $\eta$. As all the
geodesic segments $\eta_{i}$ are achronal, by the well-known argument
none of them can have a pair of conjugate points. This implies, in
turn, that any Jacobi field along any $\eta_{i}$ cannot be refocused. 
As the sequence $\{\eta_{i}\}$ converges to the generator
$\eta$, this must then imply, by continuity, that any Jacobi field
along the generator $\eta$ cannot be refocused as well. But the past 
incomplete generator $\eta$ must satisfy, by condition (3), the 
inextendibility condition, which requires that at least one Jacobi
field along $\eta$ should be refocused. In this way one obtains a
contradiction, which completes the proof.    
 
\vspace{5mm}
{\bf Acknowledgments} 

I wish to thank Professor J.K. Beem for helpful comments and
discussions. This work was supported by the Polish Committee for 
Scientific Research (KBN) through Grant No. 2 P302 095 06.  

%\newpage


\begin{thebibliography}{13}
\bibitem{p1}R. Penrose, {\em Riv. Nuovo Cimento} {\bf 1}, 252 (1969). 
\bibitem{hp}S. W. Hawking and R. Penrose, {\em Proc. R. Soc. London
Ser.} {\bf A314}, 529 (1970).  
\bibitem{kl}K. Lake, {\em Phys. Rev. Lett.} {\bf 68}, 3129 (1992).
\bibitem{st}S. L. Shapiro and S. A. Teukolsky, {\em Phys. Rev.}
{\bf D45}, 2006 (1992). 
\bibitem{ke}R. P. Kerr, {\em Phys. Rev. Lett.} {\bf 11}, 238 (1963).
\bibitem{he}S. W. Hawking and G. F. R. Ellis, {\em The Large
Scale Structure of Space-Time} (Cambridge Univ. Press, 1973). 
\bibitem{t1}F. J. Tipler, {\em Phys. Rev. Lett.} {\bf 37}, 879 (1976). 
\bibitem{r1}W. Rudnicki, {\em Phys. Lett.} {\bf A208}, 53 (1995).
\bibitem{t2}F. J. Tipler, {\em Phys. Lett.} {\bf A64}, 8 (1977).
\bibitem{tce}F. J. Tipler, C. J. S. Clarke and G. F. R. Ellis, in: {\em
General Relativity and Gravitation}, ed. A. Held (Plenum Press, 1980),
Vol. 2, p. 97.  
\bibitem{k1}A. Kr\'olak, {\em Class. Quantum Grav.} {\bf 3},
267 (1986); {\em J. Math. Phys.} {\bf 28}, 2685 (1987).  
\bibitem{ck}C. J. S. Clarke and A. Kr\'olak, {\em J. Geom.
Phys.} {\bf 2}, 127 (1985).
\bibitem{r2}W. Rudnicki, {\em Phys. Lett.} {\bf A224}, 45 (1996).  
\end{thebibliography}
\end{document}